\documentclass[doublecol]{epl2} 
\usepackage{amsmath}
\usepackage{amssymb}

\title{Impact of crowders on the morphology of bacterial chromosomes}
\shorttitle{Relative morphology of crowders and monomers} 

\author{Amit Kumar\inst{1,2}\footnote{E-mail: amit@iopb.res.in} \and Pinaki Swain\inst{3} \and Bela M. Mulder\inst{4}\footnote{E-mail: mulder@amolf.nl} \and Debasish Chaudhuri\inst{1,2}\footnote{E-mail: debc@iopb.res.in} }
\shortauthor{A. Kumar \etal}

\institute{                    
  \inst{1} Institute of Physics, Sachivalaya Marg, Bhubaneswar 751005, India\\
  \inst{2} Homi Bhabha National Institute, Anushaktinagar, Mumbai 400094, India \\
  \inst{3} Indian Institute of Technology Hyderabad, Kandi, Sangareddy 502285, Telangana, India \\
  \inst{4} Institute AMOLF, Science Park 104, 1098XG Amsterdam, the Netherlands 
}

\abstract{
Inspired by recent experiments on the effects of cytosolic crowders on the organization of bacterial chromosomes, we consider a ``feather-boa'' type model chromosome in the presence of non-additive crowders, encapsulated within a cylindrical cell. We observe spontaneous emergence of complementary helicity of the confined polymer and crowders. This feature is reproduced within a simplified effective model of the chromosome. This latter model further establishes the occurrence of longitudinal and transverse spatial segregation transitions between the chromosome and crowders upon increasing crowder size.}

\begin{document}

\maketitle

\section{Introduction}
The chromosome of \emph{E.\ coli} bacteria is constituted of a 1.6 mm long negatively supercoiled circular DNA strand and associated binding proteins. This chromosomal structure is suspended in a crowded cytosolic fluid and forms a membrane-less organelle, the so-called nucleoid, which occupies a central sub-volume of the cell~\cite{Wang2013}. The nucleoid along with the cytosol and other components of the cell is confined within the cell envelope, which in a typical wild {\em E.\ coli} cell is roughly cylindrically shaped with diameter $0.8\, \mu$m and length $2$-$4\, \mu$m. Thus the long chromosome has to compactify at least $10^3$-fold in order to fit inside this small confining volume ~\cite{Worcel1972,Fisher2013,Zimmerman2006}. Yet, this huge compaction has to be concomitant with functional organization of chromosome that facilitates, e.g.,  gene expression and  replication~\cite{VanDerValk2014}. 

Several physical and chemical processes mediate the compaction of the chromosome. Clearly, cellular confinement is a purely physical factor folding the chromosome. However, this is not the sole effect at play, as the nucleoid does not span the whole cell but occupies only about $1/4$ of the total cell volume~\cite{Zimmerman2006}. The DNA of \emph{E.\ coli} is $\sim 5\%$ negatively supercoiled, this under-twisting causing chain warps, folds and braid-like plectoneme structures to appear~\cite{Bates2005}. Also, cross-linking of different DNA segments by nucleoid associated proteins (NAPs) and loop formation by active structural modification of chromosome (SMC) proteins mediate chromosome compaction~\cite{Luijsterburg2006,VanDerValk2014,Kumar2019,Holmes2000,Zimmerman2006}. Dedicated proteins converting the looped structures to topologically associated domains (TAD) are also identified~\cite{Brocken2018,Goloborodko2016a,Song2015,VanDerValk2014}. Electron microscopy, and chromosome conformation capture (3C) techniques and their variants independently provide experimental insight into loop and contact formation in chromosomes~\cite{Postow2004, Le2013,Marbouty2014}. Finally, the depletion due to cytosolic crowders can further compress the chromosome~\cite{DeVries2012,Shendruk2015a,Jun2015}. It had therefore already been argued that supercoiled DNA may collapse into a nucleoid domain due to macromolecular compression by cytosol components~\cite{Odijk1998}, which was validated further by later experiments~\cite{Cunha2001}.  Later work showed that macro-molecular crowding and confinement together impact the chromosome size and morphology~\cite{Marenduzzo2006,Jeon2016,Jeon2017,Kang2015}. Numerical simulations indeed predicted ribosome-nucleoid spatial segregation via expulsion of ribosomes by the plectonemic DNA to both ends of the cylindrical cell~\cite{Mondal2011}.

Experiments over the last decade also revealed a large-scale helical organization of chromosomes ubiquitous in rod-shaped bacteria, displaying a definite pitch to length ratio~\cite{Berlatzky2008, Fisher2013}. The emergent size, shape and dynamics of the chromosome depend on the cellular confinement~\cite{Wu2019, Swain2019}. One mechanism proposed driving this helix formation was the bacterial actin-homologue MreB which associates with the cell wall and moves along helical trajectories~\cite{VanTeeffelen2011}. On the other hand, recent experiments revealed local out- of- phase density modulation of the chromosome and ribosomes~\cite{Gray2019}, suggesting a molecular-crowding related mechanism. It has already been shown that the helical organization produces the optimal packing of filaments in cylinders {using} a purely geometric approach~\cite{Maritan2000}. Later, it was shown that such a morphology is also entropically stabilized under thermodynamic conditions~\cite{Chaudhuri2012}. 

Here, bringing together the elements of confinement, chromosome structure and macromolecular crowding, we ask how the helical organisation of the chromosome could be impacted by the presence of cytosolic crowding agents. To that end we consider a simple ``feather-boa'' model of the chromosome as a circular backbone polymer dressed by a cloud of side-loops of equal size attached to the backbone at a regular spacing~\cite{Chaudhuri2012}. In this model, related to so-called bottle brush polymers~\cite{Heermann2011,Hsu2010b}, an effective bending rigidity emerges due to the repulsion between the side-loops. Coupled to the cylindrical confinement this leads to an emergent helical morphology~\cite{Chaudhuri2012,Chaudhuri2018, Jung2019}. We use molecular dynamics simulations, in which the viscous and poly-disperse background cytosolic fluid, thought to be also fluidized by metabolic activity~\cite{Parry2014}, is represented by a Langevin heat bath. Into this environment we introduce crowding agents at various densities and sizes. 

We show that the centre of mass of the chromosome monomers under crowding conditions remains organized in a helical fashion. Strikingly, the centre of mass of the crowders localizes to a helix complementary to the one formed by the chromosome monomers. Earlier we have shown that a more coarse-grained model, using a suitable Gaussian core repulsion replacing the side-loops, recaptures the basic morphologies of the full feather-boa chromosome model~\cite{Swain2019, Wu2019} in the absence of crowders. Here we show that the complementary helicity of crowders and monomers is reproduced within this effective model as well. We take advantage of this computationally more tractable model to more systematically examine the impact of changing crowder size on the local morphology.  This enabled us to observe and characterize a hitherto unreported transverse to longitudinal spatial segregation of monomers and crowders in the cylindrical confinement upon increasing the crowder size.

\section{Full feather-boa model}
We consider a feather-boa model of the chromosome~\cite{Chaudhuri2018} consisting of a flexible backbone of $n_b$ monomers to each of which is attached a circular chain of $n_s$ monomers. The total number of monomers in the chain thus equals $n_b (1+n_s)$ beads. The consecutive monomers in the chain are taken to be bound by a shifted harmonic potential, $ V_b = ({A}/{2}) ({\bf d}_i - \sigma {\bf u}_i)^2 $, where $ {\bf d}_i = {\bf r}_{i+1}  - {\bf r}_i$, with ${\bf r}_i$ the position of the  $i$-th bead, and ${\bf u}_i = {\bf d}_i/ \mid {\bf d}_i \mid $ the local tangent vector to the chain. The self-avoidance is modeled by  a short-ranged  Weeks-Chandler-Anderson (WCA) repulsion between non-bonded beads, $V(r_{ij}) = 4{\epsilon}[(\sigma/r_{ij})^{12}-(\sigma/r_{ij})^{6}+0.25] $ for $r_{ij} < 2^ \frac{1}{6} \sigma$ and $ \beta V(r_{ij}) = 0$ otherwise~\cite{Weeks1971}. Here $r_{ij}$ denotes the separation between $i$-th and $j$-th monomer. The length and the energy scales are set by $\sigma$, $\epsilon$, respectively, and together they set the time scale $\tau = \sigma \sqrt{{m}/{\epsilon}}$. { The mass of the particle is chosen to be $m=1$.} We use the following parameters: bond strength $A = 100\epsilon/\sigma^2$, number of backbone monomers $n_b = 200$, and number of monomers per side loop $n_s = 40$. The confining cylinder has diameter $ D = 29.5 \sigma$ and length $ L = 50.74 \sigma$. All monomers are repelled from the bounding surface by the same short-ranged interaction potential $V_{wall} = 2 \pi \epsilon [({2}/{5}) ({\sigma}/{r_{iw}})^{10} - ({\sigma}/{r_{iw}})^{4} + {3}/{5}]$, if $ r_{iw} < \sigma $ and $0$ otherwise. Here, $r_{iw}$ is the separation of a monomer from the wall. { All the parameter choices in this section match with the ones used in Ref.\cite{Chaudhuri2012} that established spontaneous emergence of chromosomal helicity in the absence of crowders.}
To study the generic impact of cytosolic elements on the chromosome organization, here, we introduce $N_c = 3000$ number of so-called non-additive crowders in the system~\cite{Dijkstra1998}. 
The crowders do not interact between themselves but repel the monomers with the WCA potential $V(r_{ij})$, and the walls with $V_{wall}(r_{iw})$. 
With the above parameters, molecular dynamics simulations are performed using the velocity-Verlet scheme in the presence of a  Langevin thermostat { characterized by an isotropic friction coefficient $\gamma=1/\tau$} fixing the temperature $T = 1.0 \epsilon /k_B$ as implemented by ESPResSo molecular dynamics package~\cite{Limbach2006}. We perform the numerical integration using step size $ \delta t = 0.005 \tau$. 

\begin{figure}[!t]
\begin{center}
\includegraphics[width=0.8\linewidth]{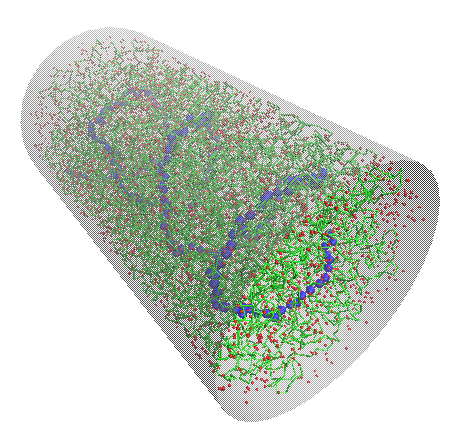}
\end{center}
\caption{A typical configuration of the feather boa chain with backbone length $l_b = 200 \sigma$ and side loop size $l_s = 40 \sigma$, in the presence of crowders. For clarity, the backbone is shown as a thick blue line. The side loops are shown in thin green line and crowders are shown by red dots.}
\label{fig:config}
\end{figure}

\begin{figure}[t]
\begin{center}
\includegraphics[width=8cm]{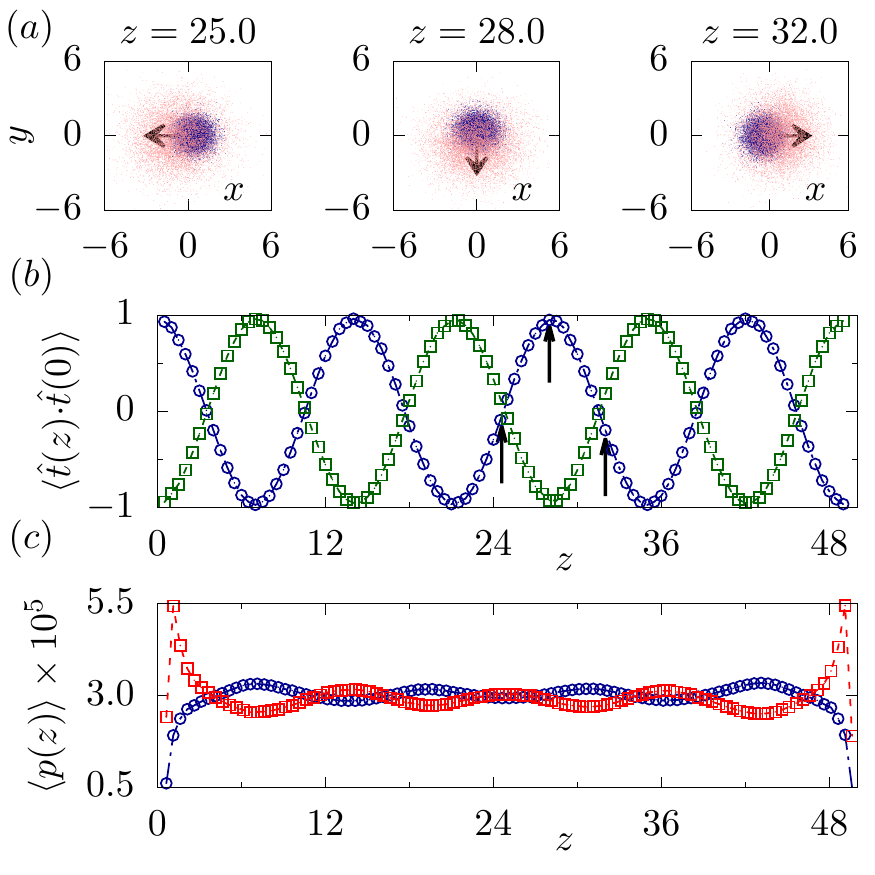}
\caption{(Color online)
($a$)~Scatter plots of centre of mass of monomers (red) and crowders (blue) in different $xy$-planes along the long axis of the confining cylinder  specified by the $z$-values within bins of size $0.5\,\sigma$ collected over $10^4$ equilibrium configurations. 
The arrow denotes relative orientation of the crowder centre of mass with respect to that of monomers. It rotates along the cell length. 
($b$)~The  correlation $\langle \hat t (z) \cdot \hat t (0) \rangle$ of the $xy$ plane- projected centre- of- mass orientation of monomers (blue $\circ$), and its cross-correlation  with that of crowder centre- of-  mass $\langle \hat t_c (z) \cdot \hat t (0) \rangle$ (green $\Box$) vary periodically with the length of the cylinder.  The periodicity $\approx 14.5 \sigma$ gives the helical pitch. The arrows on the  $\langle \hat t (z) \cdot \hat t (0) \rangle$ graph denote the longitudinal positions at which the scatter plots are shown in ($a$). ($c$)~Probability density of the monomers ($\circ$) and crowders ($\Box$) along the long axis of the cylinder.}
\label{fig:tt}
\end{center}
\end{figure}

\subsection{Complementary helical organization}

Fig.\ref{fig:config} shows a typical equilibrium configuration of the model chromosome along with the crowders.  The backbone monomers are denoted by blue beads, while the monomers belonging to side-loops are denoted by green beads. The crowders are shown by red dots. The configuration clearly shows that in the presence of crowders the backbone organises into helical morphology, a result previously obtained in the absence of the crowders~\cite{Chaudhuri2012}. As we show in the following, the crowders organize into a striking complementary helix within the cylindrical confinement of the cell.

In  Fig.\ref{fig:tt}($a$), the centres of mass of monomers (red) and crowders (blue) are shown using a scatter plot of $10^4$ equilibrium configurations separated by $5 \times 10^3 \tau$ in different bins of size $0.5\,\sigma$ at  $z=25 \sigma,\,28 \sigma,\,32 \sigma$  along the cylinder length. The arrows denote the average in-plane ($xy$) separation between the crowders and the monomers, which is seen to rotate along the cylinder length. It thus appears that the crowders preferentially localize to a helical domain complementary to the helix formed by the chromosome backbone.   

In order to quantify this complementary helicity, we compute tangent-tangent correlations~(Fig.\ref{fig:tt}($b$)\,). We first consider the relative vectors connecting the centre of mass of monomers in consecutive bins along the cell length. We obtain the correlation $\langle \hat t(z) \cdot \hat t(0) \rangle$ of the corresponding unit vectors $\hat t(z)$ projected on the transverse $xy$-plane. This is shown in Fig.\ref{fig:tt}($b$) with data denoted by blue $\circ$'s. Next, we consider $\hat t_c(z)$, relative orientation of vectors connecting the centre of mass of crowders projected onto the $xy$-plane. The cross-correlation of this orientation and that corresponding to the backbone monomers,$\langle \hat t_c(z) \cdot \hat t(0) \rangle$, are shown as the green $\Box$'s in Fig.\ref{fig:tt}($b$). This cross-correlation oscillates with the same periodicity but exactly out- of- phase with respect to the projected orientation correlation of monomer centre of mass. This clearly establishes a complementary helical organization of the chromosome and crowders with a pitch of $14.5 \sigma$.

In Fig.\ref{fig:tt}($c$) we also show the probability density profiles of the monomers and crowders. The figure shows that there are also out-of-phase density modulations of the two components along the length of the cylindrical confinement. Note the different behaviour at the cell ends: The crowders wet the two caps of the cylinder, whereas the monomer density vanishes there. Similar out- of- phase modulation of chromosome and ribosome density has been observed recently in \emph{E.\ coli} bacteria~\cite{Gray2019}.   

Together these results show that the emergent helicity of chromosomal morphology not only is robust with respect to the introduction of crowders, but moreover imprints itself on the spatial distribution of the crowders. 

\begin{figure*}[t]
\begin{center}
\includegraphics[width=16cm]{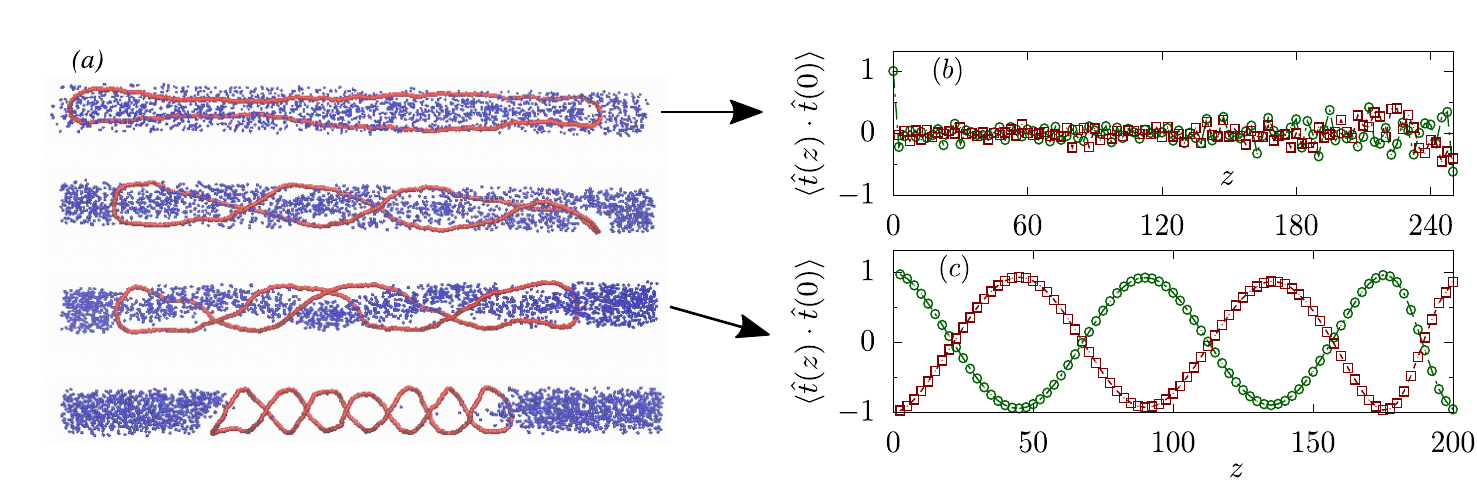}
\caption{(Color online) 
($a$)~Top to bottom: ${\Sigma_c}$ varies from  $2.0\sigma$, $3.0\sigma$, $3.5\sigma$, and $4.0\sigma$, respectively with $L_z = 12D$. Below ${\Sigma_c} = 3.0\sigma$, the model chromosome opens up with the crowders homogeneously distributed. At around ${\Sigma_c} = 3.0\sigma$, the crowders start to spatially  segregate from the chromosome, and at ${\Sigma_c} = 3.5\sigma$ we observe complimentary helicity of the chain and local crowder- density. Beyond ${\Sigma_c} = 4.0\sigma$, a clear longitudinal separation of chromosome and the crowder positions are observed.
($b$),($c$)~The structures are analyzed using the projected tangent-tangent correlation $\langle \hat t (z) \cdot \hat t (0) \rangle$ in $xy$ plane of the monomers (green $\circ$), and cross-correlation $\langle \hat t_c(z) \cdot \hat t(0)\rangle$  between monomers and crowders (brown $\Box$) are shown along the length of the cylinder.
($b$)~Orientations are uncorrelated at ${\Sigma_c} = 2.0 \sigma$. ($c$)~Complementary helicity is observed  in terms of the out- of- phase oscillations of the two correlations at ${\Sigma_c} = 3.5 \sigma$. 
}
\label{fig:confs}
\end{center}
\end{figure*}

\section{Coarse-grained model}
Having established the main features of our model feather-boa chromosome in the presence of crowders, we wish to further explore how these effects depend on the degree of crowding. In order to make this exploration computationally more tractable, we further coarse-grain our chromosome model by replacing the side-loops by an effective Gaussian core repulsion\cite{Chaudhuri2012}. The strength and range of this interaction depends on the radius of gyration of the side loops~\cite{Chaudhuri2012,Swain2019}.  The additional Gaussian core interaction between backbone monomers to incorporate effective thickening due to side loops is given by $V_{gc}(r_{ij}) = a \exp [{-r_{ij}^{2}}/{2 \Sigma^{2}}]$, with $\Sigma^{2} \sim 2 R_{g}^{2}$, where the radius of gyration of each side loop $ R_g = c n_{s}^{3/5} \sigma$ with $c=0.323$, a number obtained from separate numerical simulations~\cite{Chaudhuri2012,Bolhuis2001}. 

{  In this section we choose parameters to model the 1.6\,mm (4.6 Mbp) long circular DNA of {\em E.coli} in a $12\,\mu$m long filamentous cell grown inside a $1\,\mu$m diameter confining channel under appropriate genetic and biochemical control~\cite{Wu2019}. As we will show, it is interesting to study the impact of crowders in such a longer cell, as it provides more possibilities for the crowders to segregate spatially from the chromosome.  Choosing  $\sigma = 0.04\,\mu$m (115\,bp), we take a circular backbone of $n_b=636$ monomers and a side loop size of $n_s=62$. In these units $D=26.67\,\sigma$, yielding a value close to the $0.8\,\mu$m diameter of unconfined wild-type {\em E. coli} cells.} 

The radius of gyration of  the chosen side-loops $R_g = c n_s^{3/5}\sigma = 0.14\,D$. 
Since the loop size is much smaller than the confining diameter $D$, the effective repulsion between the side-loops can be modeled as that in bulk. The strength of effective repulsion between polymers depends on their topology. For that between long open chains, it is known to be $2\,k_B T$~\cite{Bolhuis2001}. On the other hand, the repulsion strength between circular chains varies between $2$-$6\,k_B T$~\cite{Narros2013}. To incorporate the possibility of both loops and plectoneme-like morphologies of side-loops, here we assume the intermediate value $a = 3\,k_B T$. In addition to $V_{wall}$, the repulsion between the backbone monomers with the wall also contains a soft Gaussian core contribution $V_{gc}(r_{iw})$ with width $\Sigma=R_g$ and strength $a/2$ to model repulsion from the wall due to the side-loops.
Finally, for the sake of consistency with earlier publications~\cite{Chaudhuri2012,Swain2019}, the bonds along the chain are now maintained by a finitely extensible non-linear elastic (FENE) potential~\cite{Kremer1990}  $V_F(r_{i+1,i}) = -({k}/{2}) \ln[1-(r_{i+1,i}/R)^2]$, where $k = 30$ and $R = 1.5$, in addition to the short- ranged WCA repulsion between all the particle pairs.

As in the previous section, we consider
{ non-additive crowders that do not interact between themselves but interact repulsively with the monomers and the confining walls. In addition to the WCA interaction, the crowders considered in this section, repel the chromosome with a Gaussian core  $V_{gc}^{c}(r_{ij}) = a \exp [{-r_{ij}^{2}}/{2 \Sigma_{d}^{2}}]$ that models their interaction with the side-loops, where $a$ is the same parameter as above. Also, their repulsion from the walls of the cylindrical confinement is also considered to be a combination of $V_{wall}$ and $V_{gc}(r_{iw})$, i.e., the same as that between backbone monomers and walls. To account for this larger effective crowder size, we use a smaller number of crowders $N_c=2000$.}
We perform molecular dynamics  simulations of this coarse-grained model with step-size $\delta t = 0.01\,\tau$ using the velocity-Verlet algorithm in the presence of a Langevin thermostat  keeping the temperature constant at $T= 1.0 \epsilon/k_B$~\cite{Limbach2006}. 
As a proxy for tuning the degree of crowding we choose the diameter ${\Sigma_c}$ of the crowders, as this governs to zeroth order the strength of the induced depletion interactions \cite{Asakura1958}. We then systematically explore the impact of increasing ${\Sigma_c}$ on the relative organization of the model chromosome and crowders. 

\subsection{Impact of crowder size on organization}  
Fig.\ref{fig:confs}($a$) shows how the relative organization of the model chromosome and crowders changes with increasing crowder size ${\Sigma_c}$. 
The smallest crowders get distributed homogeneously over the cylindrical confinement, and as the cylinder size of $L=12\,D$ allows it, the chromosome opens up completely. As the crowder size increases to  ${\Sigma_c} = 3\,\sigma$, the chromosome and crowders start to get spatially segregated. Crowders start to compress the chromosome into a helicoid shape. At ${\Sigma_c} = 3.5\,\sigma$, a complementary helical organization of the local crowder and monomer density can be observed in Fig.\ref{fig:confs}($a$). At even larger ${\Sigma_c}$, crowders and  the chromosome undergoes a complete longitudinal segregation in the direction parallel to the long axis of the cylindrical confinement. 

\subsection{Complementary helicity}
A complementary helical organization of the chromosome and crowders is observed in the intermediate range of crowder size ${\Sigma_c}$. It is again quantified using the tangent-tangent correlation function $\langle \hat t(z) \cdot \hat t (0) \rangle$ between monomer centre- of- mass orientations and cross-correlation $\langle \hat t_c(z) \cdot \hat t (0) \rangle$ between monomer and crowder centre- of- mass orientations along the length of the cylinder~(Fig.\ref{fig:confs}($b$)-($c$)\,). Here, the averaging is performed over $10^4$ well-separated configurations. The out- of- phase oscillation of the monomer-monomer correlation and crowder-monomer cross-correlation also illustrates the complementary helical organization of the monomers and crowders~(Fig.\ref{fig:confs}($c$)\,). Clearly, this type of organization depends on the crowder size. First of all, it requires a minimal crowder size before it appears~(Fig.\ref{fig:confs}($b$)\,). Then, at very large ${\Sigma_c}$ ($\gtrsim 4\, \sigma$) the crowders and monomers get longitudinally segregated, and the complementary helical organization disappears. 

\begin{figure}[t]
\includegraphics[width=8cm]{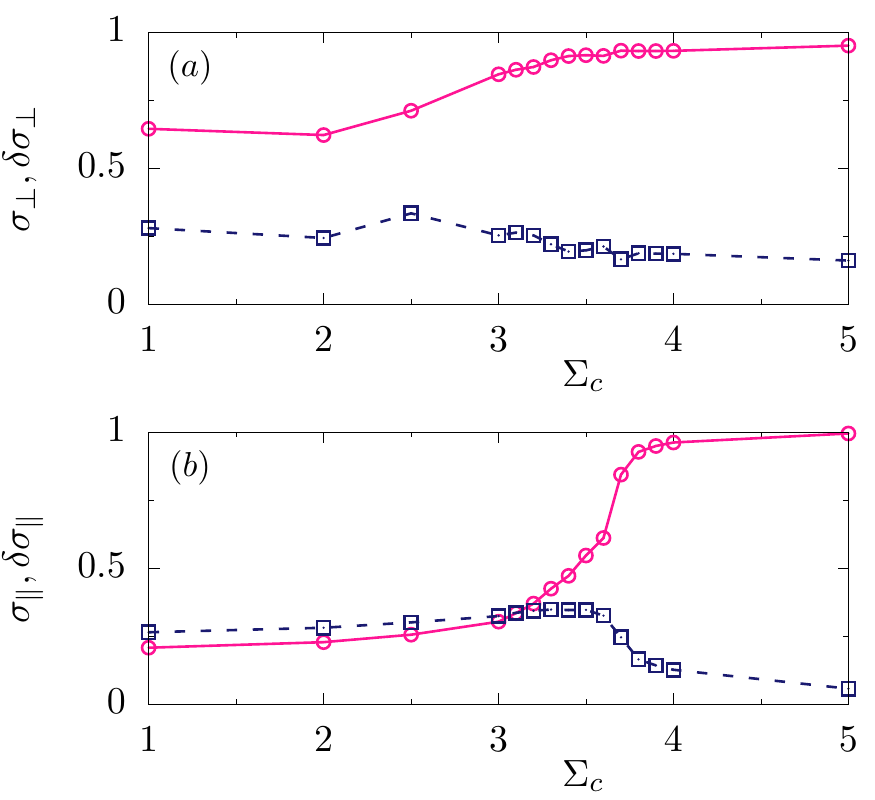}
\caption{(Color online) ($a$) Transverse spatial segregation $\sigma_\perp$ ($\circ$) and the corresponding root mean squared fluctuations $\delta \sigma_\perp$ ($\Box$) as a function of crowder size ${\Sigma_c}$. 
($b$)~Change in the longitudinal separation $\sigma_\parallel$ ($\circ$) and the corresponding root mean squared fluctuations $\delta \sigma_\parallel$ ($\Box$) with ${\Sigma_c}$.}
\label{fig:segre}
\end{figure}

\begin{figure}[t]
\includegraphics[width=8cm]{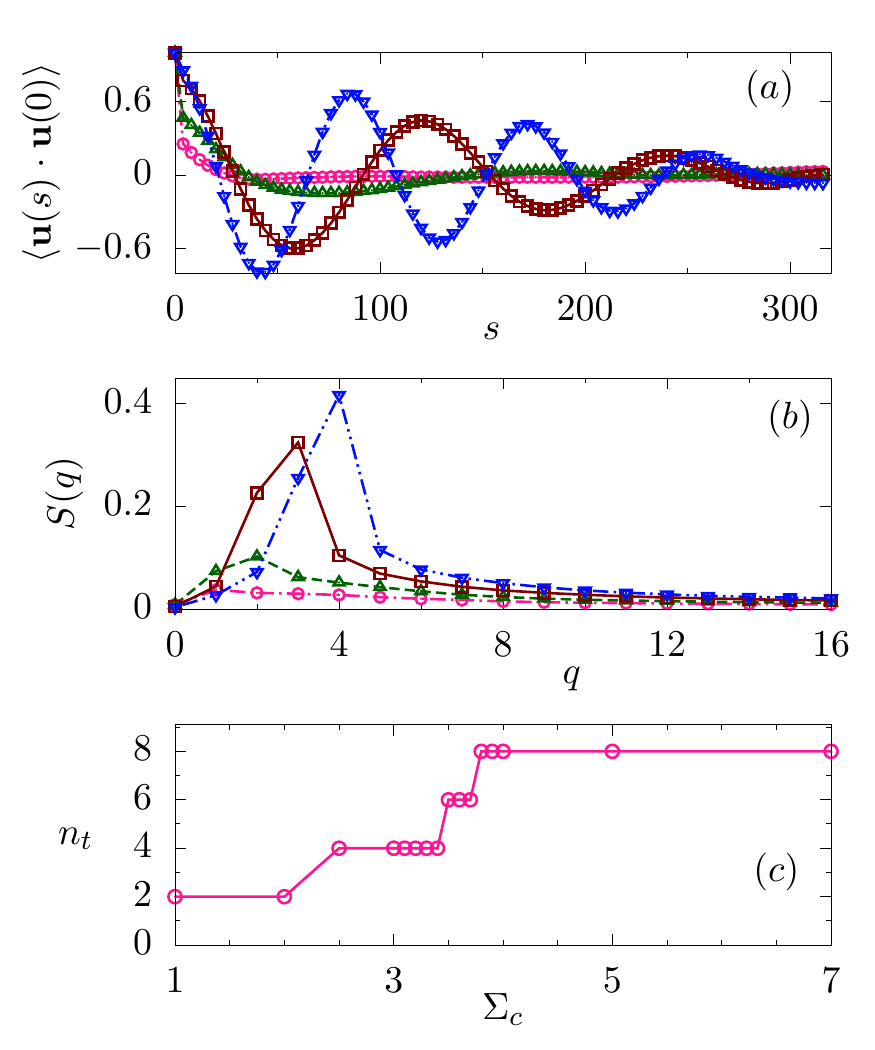}
\caption{(Color online) ($a$)~The tangent-tangent correlation  functions along the contour of the chain $s$ evaluated at ${\Sigma_c}$ = 2.0 ($\circ$), 3.0 ($\triangle$), 3.5 ($\square$) and 4.0 ($\triangledown$), respectively. ($b$)~The corresponding structure factors evaluated at the same values of  ${\Sigma_c}$, showing peaks at $q=q_p$. ($c$)~The number of turns $n_t = 2q_p$  increases with ${\Sigma_c}$. }
\label{fig:corr_2D}
\end{figure}

 \subsection{Transverse and longitudinal segregation} 
The longitudinal (transverse) spatial segregation can be quantified in terms of the following order parameter  
 $$
 \sigma_{\parallel, \perp}=\left\langle \left | \frac{\rho_m - \rho_c}{\rho_m + \rho_c} \right | \right\rangle,
 $$
 which is the average over local values measured in $120$ bins ($10$ bins) dividing the cell into longitudinal (radial) sections.  
Here, $\rho_{m,c}$ denotes the density of the monomers and crowders within each bin, respectively, normalized by their respective total numbers $N_{m,c}$. 
For the evaluation of the transverse spatial segregation, we restricted the calculations to the portion of the cylindrical confinement actually containing the chain. 
The averaging is again performed over $10^4$ well-separated configurations. Fig.\ref{fig:segre} shows the variations of $\sigma_{\parallel, \perp}$ with ${\Sigma_c}$. The corresponding root mean squared deviation $\delta \sigma_{\parallel, \perp}$ also shows an increase, albeit small, near the crossover points marking the different segregation regimes. The onset of radial segregation at ${\Sigma_c} \approx 2\,\sigma$ precedes that of the longitudinal segregation near ${\Sigma_c} \approx 3\,\sigma$. Finally, the segregation is fully completed beyond ${\Sigma_c}=4\,\sigma$.

\subsection{Change in helicity}
A change in helicity is also associated with the spatial segregation and resultant compression due to crowders upon increasing the crowder size ${\Sigma_c}$. This can be quantified in terms of the tangent-tangent correlation function $\langle {\bf u}(s) \cdot {\bf u}(0) \rangle$ along the backbone chain. Here $s$ denotes a segment of the chain along its contour.  To count the number of turns around the long axis of the cylinder, we use the radially projected tangent vectors of the backbone chain ${\bf u}(s)$.  The correlation function is plotted up to $n_b \sigma/2$, half the chain length, the longest separation along the chain~(Fig.\ref{fig:corr_2D}($a$)). 
As the filament gets into a helical shape this correlation starts to show oscillations, and the periodicity of the oscillation captures the helical pitch. The structure factors $S(q)$~(Fig.\ref{fig:corr_2D}($b$)) corresponding to the tangent correlation show peaks at $q=q_p$ which provides the total number of helical turns $n_t = 2 q_p$. The magnitude $S(q_p)$ quantifies the degree of helicity, the more pronounced helices at larger ${\Sigma_c}$ showing a correspondingly higher value of $S(q_p)$. Finally, Fig.\ref{fig:corr_2D}($c$) shows the increase in the total number of helical turns along the chain $n_t$ with the increase in crowder-size ${\Sigma_c}$. 

\section{Conclusion}
We presented a coarse-grained model of the bacterial chromosome in the form of a ``feather-boa'' model, a model of self-avoiding polymer dressed by side loops. In the presence of crowders, under cylindrical cellular confinement, we find that the monomers of the chain and the crowders arrange themselves in a complementary helical morphology. The corresponding local densities show out- of- phase oscillations along the cell length, a behavior similar to the modulation of chromosome and ribosome densities recently observed in \emph{E.\ coli}~\cite{Gray2019}.  A real cell contains proteins of various sizes impacting the chromosome organization to a various degree. A further coarse-grained chromosome model, replacing the side loops with an additional Gaussian core repulsion between the backbone monomers, was used to investigate the impact of change in crowder size on the chromosome organization. We found that the complementary helical morphology of the chromosome and crowders is retained for a range of crowder sizes. The smallest crowders permeate the whole cellular volume allowing the chromosome to completely open up in long enough cells. However, with increase in crowder size, the chromosome and crowders spatially segregate, initially in the radial direction and ultimately in the longitudinal direction. The radially segregated phase also shows complementary helicity. The number of helical turns increases as the spatial segregation proceeds with increasing crowder size.    

In an earlier study~\cite{Swain2019} we have investigated the impact of changing cell length on the chromosome organization. 
The current study establishes the impact of crowder size on the relative organization of the chromosome and molecular crowders, a topic of recent experimental interest~\cite{Yang2019}.  
The present model is able to simultaneously account both for the compression of the chromosome to a sub-volume of the cell forming a nucleoid-like membrane-less organelle, as well as the induced helicity. 
Moreover, it highlights the interesting physics arising from the subtle interplay between crowding, confinement and chromosome morphology. 

\section{Acknowledgement}
The simulations are performed on SAMKHYA, the high performance computing facility at IOP, Bhubaneswar. The work of BMM is part of the research programme of the Netherlands Organisation for Scientific Research (NWO). DC acknowledges SERB, India, for financial support through grant No. EMR/2016/001454.

\bibliographystyle{eplbib}

\end{document}